\documentclass[%
 aip,
 amsmath,amssymb,
 reprint,%
]{revtex4-1}

\usepackage{graphicx}
\usepackage{dcolumn}
\usepackage{bm}
\usepackage[mathlines]{lineno}
\usepackage{amsthm, amsfonts}
\usepackage{commath}
\usepackage[utf8]{inputenc}
\usepackage[T1]{fontenc}
\usepackage{mathptmx}
\usepackage{xcolor}
\begin{document}

\preprint{AIP/123-QED}

\title{Zooming into chaos for a fast, light and reliable cryptosystem}

\author{Jeaneth Machicao} 
 \email{machicao@usp.br}
 
\author{Odemir M. Bruno}
 \email{bruno@ifsc.usp.br}
\affiliation{ 
Scientific Computing Group. S\~{a}o Carlos Institute of Physics, University of S\~{a}o Paulo, S\~{a}o Carlos - SP, PO Box 369, 13560-970, Brazil.
}

\author{Murilo S. Baptista}
\email{murilo.baptista@abdn.ac.uk}
\affiliation{%
Institute for Complex Systems and Mathematical Biology, University of Aberdeen, AB24 3UX, Aberdeen, UK.
}%

\date{\today}

\begin{abstract}
In previous work, the $k$-logistic map [Machicao and Bruno, Chaos, {\bf 27}, 053116 (2017)] was introduced as a transformation operating in the $k$ less significant digits of the Logistic map. It exploited the map's pseudo-randomness character that is present in its less significant digits. In this work, we comprehensively analyze the dynamical and ergodic aspects of this transformation, show its applicability to generic chaotic maps or sets, and its potential impact on enabling the creation of a cryptosystem that is fast, light and reliable. 
\end{abstract}

\maketitle

\begin{quotation}
Motivated by today's huge volume of data that needs to be handled in secrecy, there is a wish to develop not only fast and light but also reliably secure cryptosystems. Chaos allows for the creation of pseudo-random numbers by low-dimensional transformations that need to be applied only a small number of times. These two properties translate into a chaos-based cryptosystem that is both fast (short running time) and light (little computational effort). The reliability of security in a chaos-based cryptosystem is sustained by the exponentially fast decay of the correlation of points in a chaotic trajectory. However, chaos is deterministic, and as such, a sufficiently long observation of the trajectory or its symbolic encoding can reveal its past and future evolution, thus breaking security. That is a known weakness of chaos in cryptography. However, this vulnerability can be compensated by another still not much explored the peculiar property of chaos. Look at the less significant digits of a chaotic trajectory and surprisingly the information content of past and future vanishes exponentially fast. What we propose here is an enhanced chaos-based cryptosystem that uses the ``deep-zoom'' transformation on the top of a chaotic map to improve the reliability of security, but without compromising on the speed and weight of the cryptosystem. We use low-dimensional chaotic maps to quickly generate numbers that have little correlation, and then we quickly (``fast'') enhance secrecy by several orders (``reliability'') with very little computational cost (``light'') by simply looking at the less significant digits of the initial chaotic trajectory. This paper demonstrates this idea with rigour, making a comprehensive ergodic characterization of this procedural strategy to create pseudo-random numbers that can sustain a fast, light and reliable cryptosystem.
\end{quotation}

\section{Introduction}

The secrecy in chaos-based cryptosystems relies on mathematical transformations that generate a trajectory whose correlation decays rapidly. The correlation of chaotic trajectories will always decay to zero after a sufficiently long time. This is due to the mixing property that allows nearby points to be quickly mapped anywhere in the transformation domain, and due to the sensibility to the initial condition chaotic transformations have. In fact, the speed of correlation decay and the sensibility to the initial conditions quantified by the Lyapunov exponent are intimately connected \cite{slipantschuk2013relation}. A chaotic system with a very large positive Lyapunov exponent is thus desirable for cryptography \cite{pisarchik2012chaotic}, since it allows for very rapid decay of correlations. Moreover, chaotic signals can be generated by low-powered, small area and simple integrated as well as analog circuits operating in very large frequency bandwidths. 

Cryptosystems need to perform heavy calculations. For example, chaos-based block ciphers \cite{fridrich1997image,farajallah2016fast,zhang2013image} such as those that encrypt images, movies and audio employ a series of complex mathematical transformations over too many bits of information. If one wants a light cryptosystem that can be run in any portable devices or that can be considered even for massive streaming, the use of real numbers with higher precision should be avoided. To improve on the performance of chaos-based cryptosystems, the underlying chaotic transformation has been discretized by considering transformations operating on an integer domain. Discretization can preserve important ergodic properties as the mixing property and the sensibility to the initial conditions, but might also create spurious periodic cycles of low-period \cite{farajallah2016fast,garasym2016new}, which result in correlations weakening the security of cryptosystems that rely on these transformations. Even chaotic transformations (such as the Bernoulli shift map) acting on real numbers with finite resolution might create spurious periodic cycles due to numerical errors. 

With recent advances, it is relatively easy today to perform numerical computation with arbitrary precision, and thus current efficient cryptosystems can rely on maps with real arithmetics of higher precision. However, any meaningful encoding of the chaotic trajectory that allows decoding, such as those used to create a pseudo-random number (PRN) generator or binary secret keys, would be strongly correlated with the most significant digits of the trajectory. To create a stream cipher based on chaos \cite{vidal2014fast}, where a binary information stream is encoded by XOR transformation to a binary secret key created by encoding a chaotic trajectory, Gerard Vidal Cassanya \cite{cassanya2017method} has proposed the use of the less significant digits of a trajectory obtained from a higher-dimensional chaotic system of ODEs to create the binary secret key. The idea of using the less significant digits of chaotic trajectories has appeared before in the work of Ref. \cite{lee2003generating}, however it was in Ref. \cite{cassanya2017method} (and other previous patent applications cited within) that less significant digits were taken by a transformation that this work claims to be optimal to support a fast, light and reliable cryptosystem.

Inspired by today huge volume of data that needs to be handled in secrecy, there is a desire to develop not only fast (quick run time) and light (little computational cost) but also reliable (highly entropic, sensitive to the initial conditions, low correlation) cryptosystems. An important aspect of a cryptosystem is its initialization. For example, one might employ a PRN to choose parameters. Secret keys, which can be created from PRNs, are also used to encrypt the information and represent a core operation in any cryptosystem. Any innovative invention that creates reliable PRNs or secret keys with optimized use of computational resources will contribute tremendously to a world that wants to communicate massive amounts of information, but securely. PRNs are not only important for secrecy in communication. It is also fundamental to the functioning of several autonomous machines, toys, and they are essential for several numerical algorithms. This work demonstrates that looking at the less significant digits of chaotic trajectories is indeed a pathway for the creation of fast, light and reliable PRNs. 

The work of Ref. \cite{machicao2017} has analyzed the dynamics and the statistical properties of the deep-zoom transformation to a chaotic trajectory, a transformation that takes up the less significant digits of a real number. This transformation applied to the Logistic map regarded as the $k$-logistic map \cite{machicao2017} was defined by the less significant digits located at $k$ digits to the right of the decimal point. It was shown that a PRN based on the $k$-logistic map has strong properties regarding statistical randomness tests DIEHARD and NIST, and thus demonstrating from a statistical perspective that the $k$-logistic map can sustain secure cryptosystems. The $k$-logistic map takes advantage of not only having trajectory points with arbitrarily large precision, and thus within principle no detectable spurious cycle, but also on hiding the information about the most significant digits, which could reveal information about the algorithm behind the generation of the PRNs. 

The interest in the present work is to understand how the deep-zoom transformation changes a particular map ergodic properties such as its space partition, density measure, Lyapunov exponent, Topological and Shannon's entropies. Our results, mostly illustrated by how the deep-zoom transformation operates into the Logistic map are valid to generic 1D chaotic maps or a set of numbers generated by any other process. The deep-zoom transformation is a complementary operation to chaos-based cryptosystem: we first quickly generate a chaotic trajectory by a low-dimensional map, and then we use the deep-zoom transformation to quickly and lightly enhance security. This is our strategy for the creation of a fast, light and reliable chaos-based cryptosystem. 

Our first result is to demonstrate that the $k$-deep-zoom ($k$-DZ) transformation to a point $x$ is mathematically equivalent to iterating for $k$ times the decimal shift map (DSM) \cite{graham1989concrete,hilborn2000chaos}. This map is well known, and it is since decades considered to be a mathematical toy model to demonstrate how a shift into the less significant digits results in strong chaos. Despite its tremendous appeal due to the nice way this map deals with decimal digits, scientists working with encryption based on nonlinear transformations have focused their attention on other more known similar maps, such as the Bernoulli shift map \cite{saito2018pseudorandom} or the Baker's map, instead of the DSM. The main difference is whereas the DSM operates by shifting the decimal numbers, Bernoulli shift and Baker's map shift the binary sequence encoding the real numbers of the trajectory. 

Then, we demonstrate that by applying the $k$-DZ transformation only once to generic chaotic trajectories, the mapped trajectories will approach a uniform invariant measure for a sufficiently large but in practice small $k$, thus requiring much less computational effort to create numbers with uniform statistics, a standard requirement for reliable PRNs. The convergence to the uniform invariant measure also dictates the convergence of the Lyapunov exponent (LE) to the Topological and Shanon's entropy of the mapped trajectories, indicating that the transformed points have achieved the largest sensibility to the initial conditions that is possible. Having a trajectory for which the level of chaos is the same as the level of entropy means that uncertainty about the past and the future is as large as one could wish for the particular chaotic map being considered. Moreover, all these quantities are linearly proportional to $k$, thus implying that randomness (higher entropy) and the sensibility to the initial conditions (large LE) can be trivially increased by the resolution with which a trajectory is observed, and not by increasing a systems dimension or by considering higher-order iterates of the map onto itself, operations that would require computational resources. 

Throughout this paper, we will show how this map amazing properties applied to any 1D chaotic systems with finite probability measure allows for a clear path to the creation of fast (quick run time, low number of iterations), light (little computational effort, low-dimension) and reliable (uniform statistics, strongly sensitive to the initial conditions, high entropy) pseudo-random numbers or symbolic secret keys, thus supporting fast, light and reliable chaos-based cryptosystems. 

\section{The $k$-deep-zoom (k-DZ) transformation and its equivalence to the Decimal shift map (DSM)}
\label{Sec:background}

Given a 1D map $f(x)$ defined in a domain $[a,b]$ and producing an orbit $\mathcal{O}(x_0)=\{x_0, x_1,\ldots, x_t\}$ generated by the initial condition $x_0$, with a given invariant density $\rho(x)$ and probability measure $\mu(x)$, such that for an interval $\epsilon \in [a,b]$ we have that $\mu(\epsilon)=\int_{x \in \epsilon}\rho(x)dx$, the $k$-DZ transformation $\phi_k(x)$ was defined in \cite{machicao2017} by 

\begin{equation}
	\phi_{k}(x) = x 10^{k} - \lfloor x 10^k \rfloor\,, 
	\label{eq:klogistic}
\end{equation}

\noindent 
where $\lfloor$ $\rfloor$ stands for the floor function. In Ref. \cite{machicao2017}, and motivated mostly for practical reasons, a parameter $L$ was considered which set the number of less significant digits for the function $\phi_{k}(x)$. In here, we drop this definition, and assume that $L \rightarrow \infty$, or is a large number. 

This map can be analogously described by

\begin{eqnarray}
	\phi_{k}(x) = 10^k(x, \mod{10^{-k}}) \label{kDZ-1} \,.\\
	 \phi_{k}(x) = 10^k x, \mod{1} \,. \label{kDZ-2} 
\end{eqnarray}
 
The DSM map is defined by 

\begin{equation}
	D(x) = 10 x, \mod{1}, 
	\label{DSM}
\end{equation}
\noindent
and its $k$-folded version (the $k$-th iteration of $D$) which we represent by $D^k$ is basically
\begin{equation}
	D^k(x) = 10^k x, \mod{1}, 
	\label{DSM-DZ}
\end{equation}
\noindent
which is exactly equal to Eq. \eqref{kDZ-2}. Thus, the $k$-fold DSM map is mathematically equal to the $k$-DZ transformation. 

To illustrate the action of the DZ transformation, given the value $x=0.3923481$, then $\phi_{k=2}(x)= 0.23481.$

\subsection{The $k$-DZ transformation and others maps in literature}

The idea of using a cryptosystem based on mod transformations that extract the less significant digits of real numbers generated by chaotic systems was to the best of our knowledge first proposed in Ref.
\cite{lee2003generating}. Given a real number $x_n$ generated by a chaotic system (discrete or continuous), this work has proposed to cipher $x$ by 
\begin{equation}
R_n \equiv Ax_n, \mbox{\ mod S},
\label{eq:lee}
\end{equation}
with $A$ and $S$ representing arbitrary constants.

Equation \eqref{DSM} can be seen as a particular case of Eq. \eqref{eq:lee}, but not of the Eq. \eqref{kDZ-2} because the k-DZ transformation introduces the extra parameter $k$.

Our work shows when this parameter can generate suitable PRNs. However, in the work of Ref. \cite{lee2003generating}, only the case for $A=10^7$ and $S=256$ was studied, and without the rigour and deepness presented in the present work to study the ergodic and dynamic manifestations of these transformations. 
The choice of $S=256$, which turns the map of Eq. \eqref{eq:lee} not equivalent to the DSM map, was made to organize the number $R_n$ into a two-dimensional gray-scale image for further processing. This choice, however, is not optimal for the security of the encryption, measured in terms of the entropy and sensibility to the initial conditions. The optimal choice, demonstrated further, is obtained for $S=1$, as in Eqs. \eqref{kDZ-2} or \eqref{DSM-DZ}. The choice made of $A$ as an arbitrary constant is also per se not always optimal to extract the less significant digits, unless this arbitrary constant is of the form $A=10^k$, as in Eq. \eqref{kDZ-2}.

\section{The Logistic map}
The logistic map has been extensively studied over the past years~\cite{MayChaos}. Since it is a well know system and that produces typical chaotic behaviour \cite{OttChaosBook}, we focus the application of the $k$-DZ transformation to trajectory points being iterated by the logistic map, which we refer as the $k$-logistic map, adopting previously defined terminology. It is described by 
\begin{equation}
f(x_{t+1}) = b x_t(1 - x_t) \,,
\label{eq:logisticmap}
\end{equation}
\noindent where $x_t \in [0,1]$. In Eq. \eqref{eq:logisticmap}, $x_t \in \Re$, and as such each trajectory point is assumed to have infinite precision. However, in practice, $x_t$ has finite precision, but this does not prevent one from solving Eq. \eqref{eq:logisticmap} by numerical means. The shadowing lemma \cite{grebogi1990shadowing} guarantees that numerical solutions of this map are stable even if trajectory points have finite resolution, in the sense that the numerical trajectory will remain close to a true trajectory for a very long time, this time depending on the resolution of the trajectory considered.

\section{Analysis}

\subsection{Phase space, partition, and topological entropy}
 
\begin{figure*}[!ht]
\centering
{\includegraphics[scale=0.43]{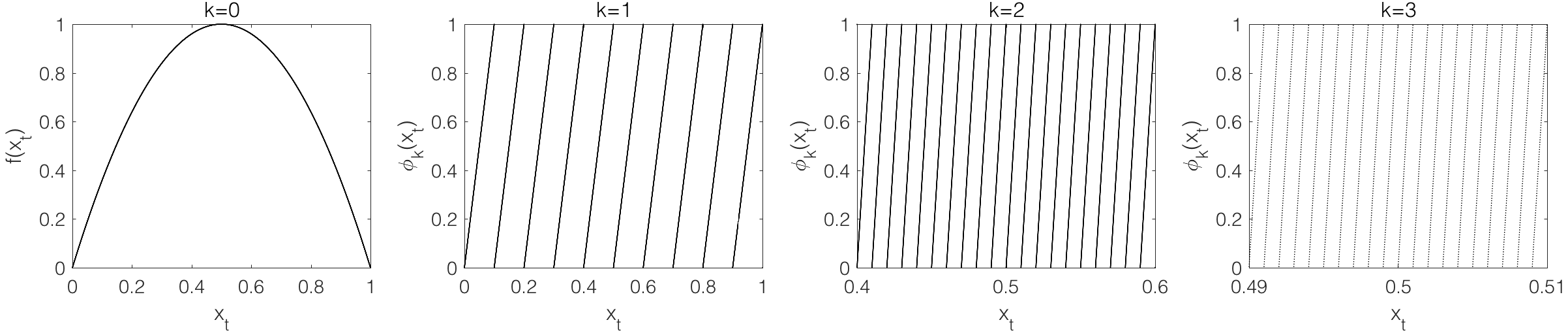}}
\caption{The $k$-DZ transformation applied to the trajectory points of the Logistic map. From left to right panels, for $k=0$, $k=1$, $k=2$, and $k=3$ by using Eq. \eqref{kDZ-2}.}
\label{figure-kDZ}
\end{figure*}

Equation \eqref{kDZ-1} is useful because it provides the key to calculate the location of the partition points, where the map becomes discontinuous. The points of discontinuities 
happen at the boundaries created by the mod function, so at multiples of $10^{-k}$, more specifically at $m10^{-k}$, for $m \in \mathbb{N}$ and $m \leq 10^k$. There will be then $10^k$ discontinuous intervals. Figure ~\ref{figure-kDZ} show the original Logistic map with $b$=4 ($k=0$, left panel), and the corresponding $k$-DZ transformation for $k=1$ (second panel to the right), $k=2$ (third panel to the right), and $k=3$ (right-most panel). 

So, the set containing the points $x^*$ where discontinuities appear can be obtained by solving the following equation 

\begin{equation}
x^*(m)= m 10^{-k}. 
\label{eq:partitions}
\end{equation}

We can define a topological entropy of the $k$-DZ transformation, which is an upper bound for the Shannon-entropy, by the Boltzmann entropy of gas measuring the entropy of it in terms of the number of observable states. Here, we can define the states as being the fall of a trajectory point into an interval within the partition provided by Eq. \eqref{eq:partitions}. Regardless of the value of $b$, and actually regardless of which kind of 1D chaotic map is used, this number is given by the number of partition points of the $k$-DZ transformation and it is equal to $10^k$, which is also the number of possible symbolic sequences that the $k$-logistic map produces. It is given by 

\begin{equation}
H_T = k \ln{(10)}. 
\label{eq:topological-entropy}
\end{equation}

It is useful to compare this result with the topological entropy of the original Logistic map, defined in terms of the number of subintervals in its generating Markov partition, and equal to $2^o$, where $o \in \mathbb{N}$ is the order of the partition representing the resolution of the subintervals composing the partition (measuring $~2^{-o}$ in length). That results in that $H_T=o\ln{(2)}$. Here we see an advantage of the use of the $k$-logistic map to produce efficient pseudo-random numbers in a light fashion, so without requiring too expensive computational resources. Assuming $o$ and $k$ to be of the same order, the topological entropy of the $k$-logistic map be $\ln{(10)}/\ln{(2)}$ larger than that of the Logistic map. 

It is also worthwhile to compare the result in Eq. \eqref{eq:topological-entropy} with the topological entropy, $H_T^{(S)}$, obtained by applying Eq. \eqref{eq:lee}. Defining $A=10^k$, we obtain that the topological entropy equals $H_T^{(S)} = k \ln{(10)} - \ln{(S)}$ for Eq. (\ref{eq:lee}). Thus, the entropy achieved in Eq. \eqref{eq:topological-entropy} for the $k$-DZ transformation in Eq. \eqref{kDZ-2} can only be achieved by applying Eq. \eqref{eq:lee} to that same chaotic set if $S=1$.

\subsection{$k$-logistic map probability density}
 
\begin{figure*}[!ht]
\centering
{\includegraphics[scale=0.8]{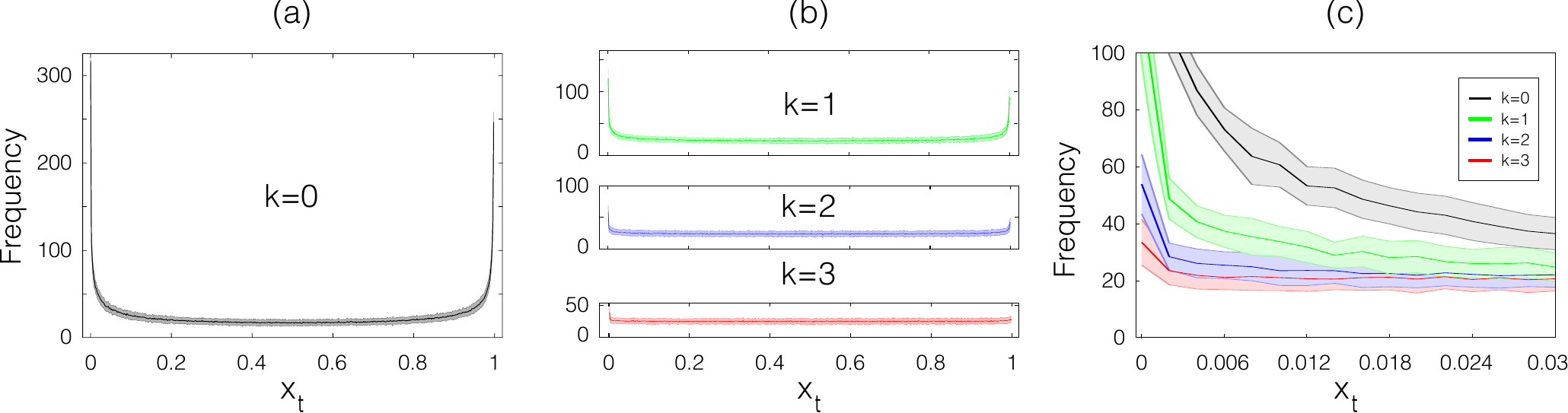}}
\caption{Frequency distribution curves for a) the original logistic map, b) the $k-$-logistic map with $k=1$, $k=2$ and $k=3$ and parameter $b=4$. The horizontal axis shows the $x_t \in [0, 1]$ (500 bins) and the vertical axis shows the frequency of the $10^4$ values discarding the first $10^3$ transient values. The curves represent the mean and standard deviation (shaded error bar) for sequences generated over 100 random initial conditions. c) The inset plot depicts a zoom on the windows $x \in [0, 0.03]$ for these 4 plots.}
\label{fig:histogramas}
\end{figure*}

One of the most important characteristics of a good PRN generator is that successive output values of it, say $u_0,u_1,u_2,\ldots$ are independent random variables from the uniform distribution over the interval [0, 1]. It was shown in ~\cite{machicao2017} that as $k$ increases the probability distribution of the map becomes more and more uniform. This is reproduced in Fig.~\ref{fig:histogramas}, in terms of the histogram (frequency) analysis. As can be seen in this figure, for $k=0$ (a) this distribution is not uniform as it is to be expected from the Logistic map with $b$=4, with a high probability of finding points close to 1 and 0. As $k$ grows with $k=1$, $k=2$, and $k=3$, the distribution tends to become increasingly uniform, as can be observed in the Figure~\ref{fig:histogramas}(b). In (c) we show a magnification of (b) for the region close to zero.

\subsection{The natural invariant measure of the $k$-DZ transformation and its Shannon's entropy}

To calculate the asymptotic Lyapunov exponent of the $k$-DZ transformation, which is independent on the choice of the chaotic map, we notice that the $k$-DZ is piecewise linear, wherein each partition sub-interval the map has a constant derivative function. Arranging the values of $x^*(i)$ in Eq. (\ref{eq:partitions}) in a ranking of crescent order and indicating it by, i.e., $x^*(m) \equiv x^*_{i}$ such that $x^*_{i+1}>x^*_{i}$, each partition subinterval comprises the interval 

\begin{equation}
 d_{i} = [x^*_{i}, x^*_{i+1}[, 
\label{eq:subpartitions}
\end{equation}
\noindent
for $i \in \mathbb{N}$ and $i=[0,1, \ldots, 10^k-1]$.

The derivative of the piecewise-linear map for each sub-interval $d_i$ can be calculated by 
\begin{equation}
\omega_i = |d_i|^{-1}=10^{-k}, 
\label{derivative2}
\end{equation}
\noindent
since $\phi_k(d_i) = 1$, where $| d_i |$ represents the length of the sub-partition $d_i$. 

The evolution of an arbitrary initial probability measure to a 1D nonlinear transformation is dictated by the Perron-Frobenious operator. For piecewise linear systems, the Perron-Frobenious operator can be cast in terms of a linear system of equations operating in each subinterval of the map partition. The $k$-DZ transformation takes as the initial measure generated by the nonlinear Logistic map and then applies $k$ times the $DSM$. If we assume that the measure in each subinterval of the $k$-DZ is uniform (which initially will be not) and we represent it by the component $[{\mathbf \mu}]_i$ of the vector ${\mathbf \mu}$ ($i=\{1, \ldots,n\}$) with $n= 10^{k}$, and we define the density in each interval as given by 

\begin{equation}
 \rho_i = \frac{\mu_i}{d_i} 
 \label{eq:density}
 \end{equation} 
 
\noindent
an equation for the evolution of the non-normalized density at iteration $t$ can be obtained \cite{AlligoodBookChaos}.

\begin{equation}
{\mathcal Z}{\mathbf \rho^{\prime}}^t = {\mathbf \rho^{\prime}}^{t+1}, 
\label{eq:linear-operator}
\end{equation}

\noindent
where the square matrix ${\mathcal Z}$ with component $[{\mathcal Z}]_{ij}$ is the reciprocal of the absolute value of the slope of the map taking the measure from the interval $j$ to the interval $i$ and can be defined by a matrix with equal rows as 
\[
 {\mathcal Z} =
 \left[ {\begin{array}{ccccc}
 \omega_0^{-1} & \omega_1^{-1} & \ldots & \omega_{n-1}^{-1} & \omega_n^{-1} \\
 \omega_0^{-1} & \omega_1^{-1} & \ldots & \omega_{n-1}^{-1} & \omega_n^{-1} \\
\multicolumn{5}{c}{\dotfill} \\
\omega_0^{-1} & \omega_1^{-1} & \ldots & \omega_{n-1}^{-1} & \omega_n^{-1} \\
 \end{array} } \right].
\]

Equation \eqref{eq:linear-operator}, representing how the measure evolves concerning only 10$^{k}$ intervals are valid if the measure and the density is uniform for every sub-partition $d_i$ since it has been derived from the continuous Perron-Frobenious operator integrated over intervals where the measure was assumed to be constant. If the initial measure is not uniform for each subpartition interval, as it is the case since the initial measure was generated by the logistic map, we either should consider the continuous operator (effectively described by an infinite dimension matrix) or alternatively, we can adopt a much simpler strategy. We take Eq. \eqref{eq:linear-operator} and study it in the limit, when $t \rightarrow \infty$. 

Defining the vector $\mathbf{d}=\{d_1,d_2,\ldots,d_n\}$ and the diagonal matrix $D=\mathbb{I} \mathbf{d}$, the element $[]_{ij}$ of the matrix $D{\mathcal Z}D^{-1}$ represents the percentage of the measure in the interval $d_j$ that goes to the interval $d_i$. The matrix ${\mathcal Z}$ has equal rows because the piecewise equivalent of the $k$-logistic map takes measure from each interval to all others with the same proportion in each of the intervals $d_j$. 

The equilibrium point of Eq. \eqref{eq:linear-operator} is obtained when 
\begin{equation}
{\mathcal Z}{\mathbf \rho^*} = {\mathbf \rho^*}, 
\label{linear-operator1}
\end{equation}
\noindent
which means that the time invariant density is a normalized eigenvector of ${\mathcal Z}$.

The matrix ${\mathcal Z}$ is a stochastic matrix, since it is a non-negative matrix and the sum of all elements in a row totals 1. This is easy to see since
\begin{equation}
\sum_i \omega^{-1}_i = \sum_i d_i = 1. 
\end{equation}

The Perron-Frobenious theorem guarantees that a square stochastic matrix has a unique dominant real unitary eigenvalue, with all other eigenvalues smaller than 1. This means that the density of the $k$-DZ transformation in the limit of $k \rightarrow \infty$ is natural (it is unique), regardless of the initial probability measure that is fed into the $k$-DZ transformation. The natural density can be recovered by proper normalization dividing ${\mathbf \rho^*}$ by $\sum_i [{\mathbf \rho^*}]_id_i$ so that the physical natural density in each interval is given by 

\begin{equation}
[\rho]_i = \frac{{[\mathbf \rho^*]_i}}{\sum_i [{\mathbf \rho^*}]_id_i}. 
\end{equation} 

\noindent
This is to guarantee that the density produces the natural measure by Eq. \eqref{eq:density}.

It is also easy to see that the unique unitary eigenvalue has associated to it a uniform eigenvector with all components equal to a constant value $c$: ${\mathbf \rho}^* = [c\, c\, c\, \ldots, c]^T$, so, the piecewise $k$-DZ transformation has a uniform density given by 

\begin{equation}
[\rho]_i = \frac{{c}}{\sum_i [c]d_i}=1. 
\end{equation}

\noindent
This leads us to an invariant natural measure in each interval that equals the Lebesgue measure of the interval, and thus 

\begin{equation}
\mu_i = d_i=10^{-k}. 
\label{natural-measure}
\end{equation}

\noindent
So, for sufficiently large $k$, it is to be expected that the $k$-logistic map will have a uniform natural invariant density, although the density of the Logistic map is not uniform for each interval. In practice, this sufficiently large number is around $k$=4, when this map generates PRNs with all the good statistical characteristics for security \cite{machicao2017}. Being invariant means that any initial probability measure will eventually evolve to the same invariant measure. Thus, the reliability of the security for the PRNs generated by the $k$-DZ transformation is substantially more dependable on the properties of the DSM, than on the statistical properties of the chaotic set of points being iterated by the $k$-DZ transformation, or also on the chaotic map considered to initially generate the chaotic trajectory to be fed into the $k$-DZ transformation. Since the invariant measure of the $k$-logistic map is constant (for sufficiently large $k$), this means that any encoding supported by the partition defined in Eq. \eqref{eq:partitions} will produce equiprobable symbols, this rendering cryptoanalysis based on frequency statistics to be inappropriate. 

The asymptotic Shannon's entropy of the $k$-DZ transformation is therefore equal to the Topological entropy: 

\begin{equation}
H_{S} = - \sum_{i=1}^{n} \mu_i \ln{\mu_i} = - \sum_{i=1}^{n} d_i \ln{d_i} = H_T
\end{equation}

\subsection{The Lyapunov exponent of the $k$-DZ transformation}

The Lyapunov exponent (LE) of the $k$-DZ transformation can always be calculated regardless of the chaotic map being used as the generator of the initial measure. This is so because the map is piecewise linear with constant derivative everywhere (except the partition points). The Lyapunov exponent can be calculated by 

\begin{equation}
\lambda = \int \ln{\left( \abs{ \frac{d\phi_{k}(x)}{dx} } \right)} d \mu 
\label{eq:lyapunov} 
\end{equation}

\noindent where $d\mu=\rho(x)\, dx$, represents the invariant measure of the $k$-DZ transformation. 

The chaotic map has its own domain of validity. This domain must be normalized to fit within the domain of the $k$-DZ transformation. For the Logistic map, the domain is $[0,1]$, the same as the domain of the $k$-DZ transformation. Therefore, its LE is equal to 

\begin{equation}
\lambda = \int_0^1 \ln{\left( \abs{ 10^k } \right)} dx = k\ln{(10)} = H_T. 
\label{lyapunov_HT} 
\end{equation}
\noindent

So, we see that for a sufficiently large $k$, the $k$-DZ transformation produces a LE that approaches the topological entropy which is also equal to Shannon's entropy. A light cryptosystem that does not require much computational effort demands the use of transformations that can be as entropic as possible and with the largest as possible sensibility to the initial conditions (which implies in a quick decay of correlation).

When compared the LE of the $k$-DZ transformation in Eq. (\ref{kDZ-2} (result in Eq. (\ref{eq:lyapunov})) with the LE of the transformation Eq. \eqref{eq:lee} (proposed in Ref. \cite{lee2003generating}), 

assuming $A=10^k$, we notice that Eq. (\ref{eq:lee}) can be equivalently written as 
\begin{equation}
R_n=S\left( \frac{10^k}{S}x_n, \mbox{\ mod 1} \right)\,,
\label{eq:lee2}
\end{equation}

\noindent which can be rewritten as 

\begin{equation}
R_n=S\Phi(x_n), 
\label{eq:lee3}
\end{equation}
where $\Phi(x_n)=\frac{10^k}{S}x_n, \mbox{\ mod 1}$. Noticing that the 
LE of the function $\Phi(x_n)$ is the same as the one obtained if 
$\Phi(x_n)$ is multiplied by a constant, then the LE of Eq. (\ref{eq:lee}) is equal to 
\begin{equation}
\lambda^{(S)}=\ln{\left( \frac{10^k}{S} \right)}=H_T^{(S)}.
\label{LE:lee}
\end{equation}
Thus, the LE of Eq. \eqref{eq:lee} is only equal to the one of Eq. \eqref{kDZ-2}, if $S=1$. In the result of Eq. \eqref{LE:lee}, we have assumed that the speed of convergence of the probability density measure \cite{boyd2009fastest} of Eq. \eqref{eq:lee} is the same as the one of Eq. (\ref{kDZ-2}). This is to be expected, since the second largest eigenvalue of the matrix ${\mathcal Z}$ regulating the evolution of the density measure for Eq. \eqref{kDZ-2} is the same as the one for this matrix regulating the evolution of the density measure for Eq. \eqref{eq:lee}, and both are equal to zero.

\subsection{Enhancement of sensibility to the initial conditions of the $k$-logistic map}

\begin{figure*}[!hbtp]
\centering
{\includegraphics[scale=0.43]{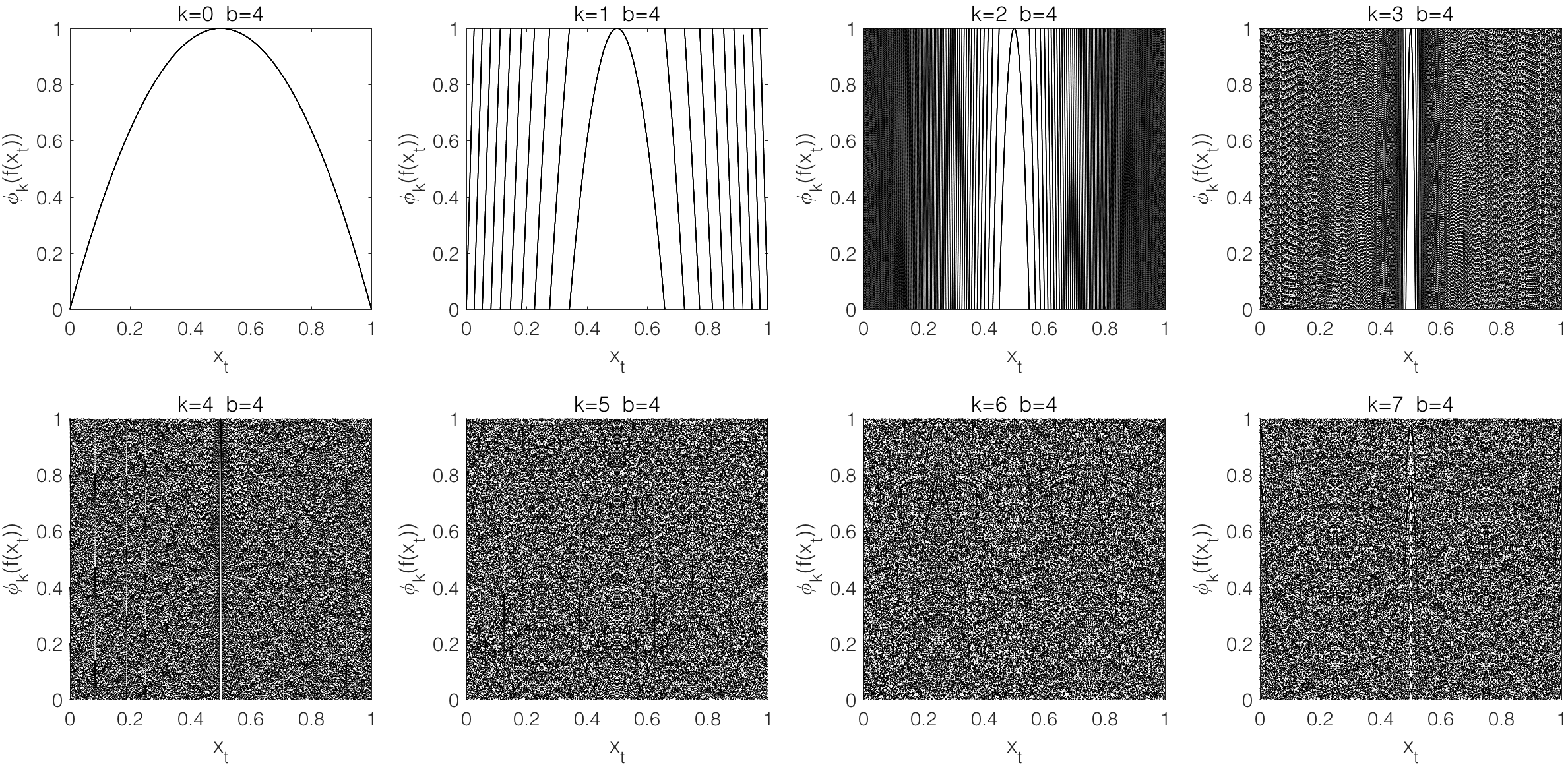}}
\caption{The $k$-logistic map state space. From left to right panels are shown $k=0$, $k=1$, $\ldots$, $k=7$ using parameter $b = 4 $. The horizontal and vertical axes show the state space of $x^k_t$ against $\phi_k(f(x_t))$. Each orbit contains $10^5$ points starting from random initial conditions.}
\label{figure-kDZ-logistic}
\end{figure*}

\begin{figure*}[!ht]
\centering
{\includegraphics[scale=0.4]{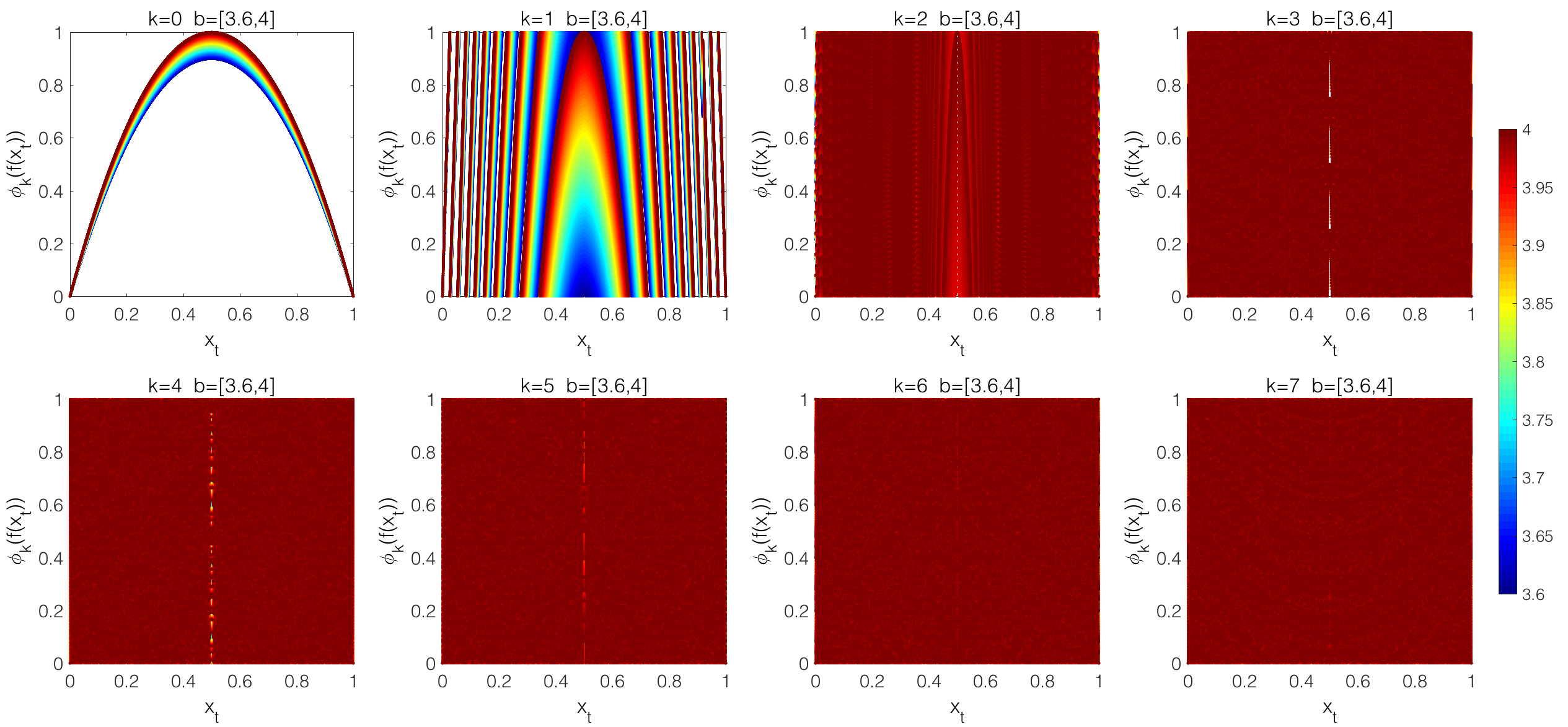}}
\caption{The $k$-logistic map state space. From left to right panels are shown $k=0$, $k=1$, $\ldots$, $k=7$ on region $b \in [3.6, 4]$. The horizontal and vertical axes show the state space of $x^k_t$ against $\phi_k(f(x_t))$. Each orbit contains $5\times 10^3$ points starting from random initial conditions.}
\label{figure-kDZ-logistic-colored}
\end{figure*}

The LE of the $k$-DZ transformation does not depend on the choice of the chaotic map generating the measure. It is nevertheless interesting to understand how much chaos is enhanced by the application of the $k$-DZ transformation into a chaotic map. Considering this chaotic map to be the logistic map (Eq. \eqref{eq:logisticmap}), we then want to understand how much chaos is enhanced if the DZ transformation with $k=1$ is applied not to the trajectory points generated by the logistic map, but to the map itself. So, we calculate the Lyapunov exponent of the map $\phi_{k}(f(x_t))$, whose state space ($\phi_{k}(f(x_t)) \times x_t$) is shown in Fig. \ref{figure-kDZ-logistic}. Additionally, Fig. \ref{figure-kDZ-logistic-colored} show a colored version of this previous picture for parameters in region $b \in [3.6, 4]$.

This map is described by 

\begin{equation}
	\phi_{k}(f(x)) = 10^k(f(x_t), \mod{10^{-k}})\,. 
	\label{eq:1DZ-1} 
\end{equation}

Its LE can be calculated by 

\begin{equation}
\lambda = \int \ln{\left( \abs{ \frac{d\phi_{k}(f(x))}{dx}} \right)} d\mu 
\label{eq:lyapunov2} 
\end{equation}
\noindent
where $d\mu=\rho(x)\, dx$ now represents the measure of the Logistic map. 

The first derivative of the map in Eq. \eqref{eq:1DZ-1} is

\begin{equation}
\frac{d\phi_{k}(x)}{dx} = 10^k b (1-2x), 
\label{eq:derivative}
\end{equation}

\noindent
whereas its density for $b=4$ is given by 

\begin{equation}
\rho(x)=\frac{\pi^{-1}}{[x(1-x)]^{1/2}}
\label{eq:analytic-density}
\end{equation}

Placing Eqs. \eqref{eq:derivative} and \eqref{eq:analytic-density} in Eq. \eqref{eq:lyapunov2} and integrating over the map domain ($x \in [0,1]$), we obtain that 

\begin{equation}
\lambda = \int_0^1 \frac{k \ln{10} + \ln{4} +\ln{|(1-2x)|}}{\pi \sqrt{x(1-x)}} dx = k \ln(10) + \ln(2), 
\label{eq:lyapunov1} 
\end{equation}
\noindent
since 
\[
\int_0^1 \frac{\ln{|(1-2x)|}}{\pi \sqrt{x(1-x)}} dx = \ln{2}.
\]

So, the first thing to notice is that the LE of the map in Eq. \eqref{eq:1DZ-1} is equal to the LE of the $1$-logistic map plus the LE of the original logistic map for $b=4$ (which is equal to $\ln{(2)}$). This tells us that when creating a cryptosystem based on a chaotic map, more entropy and sensibility to the initial conditions can be achieved by a simple inspection to the least $k$ significant digits, instead of doing more iterations in the chaotic map generating the initial chaotic sequence. 

This analysis can be easily extended to the logistic map operating under any parameter $b$ that produces chaotic motion. 

The Lyapunov exponent of the map in Eq. \eqref{eq:1DZ-1} can be calculated using the time approach by 
\begin{equation}
\label{numerical-estimation1}
\lambda(b) = \lim_{T\to\infty}\frac{1}{T}\sum_{t=1}^{T} \ln \abs{10^k b (1-2x_i)}, 
\end{equation}

\noindent
which lead us to 
\begin{equation}
\label{numerical-estimation2}
\lambda(b) = k \ln{10} + \lim_{T\to\infty}\frac{1}{T}\sum_{t=1}^{T} \ln \abs{b(1-2x_i)}, 
\end{equation}

\noindent
and finally to 
\begin{equation}
\label{numerical-estimation3}
\lambda(b) = k \ln{10} + \lambda_0(b), 
\end{equation}

\noindent
where $\lambda_0(b)$ is just the Lyapunov exponent of the Logistic map for the parameter $b$. Thus, here it is obvious that the gain for sensibility to the initial conditions is trivially achieved by just choosing a sufficiently large $k$.

\section{Pseudo-random numbers and symbolic secret keys}
\label{PRNs}

Once the output of the $k$-DZ transformation $\phi_k(x)$ generates real points in the unit interval, these values can be considered directly as a pseudo-random number that can be re-scaled as required. The security analysis of the so-called $k$-logistic map PRN was analysed in \cite{machicao2017}, showing high-quality pseudo-random numbers for $k\geq 4$ through statistical randomness tests such as DIEHARD \cite{DIEHARD} and NIST \cite{NIST-PRNG}.

Another strategy to generate PRNs is by means of the symbolic representation of the trajectory of the $k$-DZ transformation. 
Thus, a partition that is not the natural partition of the $k$-DZ transformation needs to be considered. This natural partition is given by $\bm{d}$ whose borders are defined by Eq. \eqref{eq:partitions}. Then, for a given $k$, there will be $10^k$ symbols for the natural partition. The point $\phi_k(x_i) \in [d_i,d_{i+1}]$ is encoded by the $i$-th symbol of the alphabet ($i=\{0,1,\ldots,10^k-1\}$), represented by $s_i$. 
A transformed trajectory of length $L$ represented by $\{\phi_k(x_1), \phi_k(x_2), \ldots, \phi_k(x_L)\}$ will have the symbolic representation $\bm{s} = \{s_1, s_2, \ldots, s_L\}$, where $s_i \in [0,10^k-1]$. The vector $\bm{s}$ fully represents the information about the location of the points $x_i$ being mapped (within the resolution of the partition cells), and therefore should be avoided for the creation of the secret key. The partition to create a secret key should have a minimal number of intervals, for example a binary partition where $\phi_k(x_i) < 0.5$ is encoded by `0' and $\phi_k(x_i) \geq  0.5$ is encoded by `1'. In this way,  points within $x_i \in [0,1]$ will be encoded with equal probabilities for `0' and `1'.

\section{Conclusions}

Cryptography relies on the application of several transformations to eliminate all existing correlations between the message and its ciphered version. A preliminary requirement for achieving this relies on the use of highly entropic and non-correlated pseudo-random numbers. The sensitivity to the initial conditions property chaotic systems have is the key to this goal. The interest today is to be able to accomplish such a task for reliable encryption but by relying on transformations that require little computational effort (light) and quick running time (fast). 

In this work, we characterize the properties of the so-called the $k$-Deep Zoom ($k$-DZ) to support reliable cryptosystems that uses pseudo-random numbers or secret keys that were created fast and lightly. Besides the Decimal Shift Map (DSM) is not conceptually equivalent to the k-DZ, we show that the k-DZ is mathematically equivalent to the DSM map iterated k times. More than that, we show that the $k$-fold DSM can be rewritten into a form completely equivalent to the $k$-DZ transformation. So, all the good properties of the DSM map such as uniform statistics, high entropy, and sensibility to the initial conditions are inherited by the $k$-DZ. There is a semantic difference between both maps. Whereas the $k$-DZ transformation effectively represents an algorithm that simply extracts the less significant digits of a real number, the DSM is a map that transforms a point into another point. This semantic interpretation of the DZ-transformation can be in the future exploited for the creation of dedicated electronic chips operating at the hardware level that only work with less significant digits, thus potentially bringing the encryption process to the physical level. We show that the entropy and the Lyapunov exponent is linearly proportional to $k$. This means that the trivial and light task of peeking onto the sequence of less significant digits positioned $k$ digits to the decimal floating-point is sufficient to drastically increase the entropy and therefore the uncertainty past and future numbers, at a minimal computational cost. 

Several of the properties of the $k$-DZ transformation depend only on the map itself, not on the chaotic system being considered as the generator of the original trajectory being encoded, or any other set of numbers being generated by any other process (e.g. stochastic processes). Thus, one might wonder why to use the $k$-DZ transformation into a chaotic set of numbers after all? The reason is that chaotic trajectories have several advantages. They are easy to be generated and do not require the use of higher-dimensional systems, in both digital or analog domains, they require less algorithmic complexity, less-power electronics, less CPU dedication and can be generated at impressive large bandwidths. Chaos, however, is deterministic and correlation does decay quickly, but not as quickly as one would wish. The additional application of the $k$-DZ transformations to chaotic trajectories fast and lightly enhances the already existing wished properties of chaos to cryptography. A transformation that optimizes essential ingredients to a secure cryptosystem, but with minimal computational effort. 

Our strategy to create pseudo-random numbers or secret keys requires the use of a chaotic system whose simulated trajectory is guaranteed to be chaotic for a long period, and that can be additionally generated using minimal computational resources. For this reason, the Logistic map is a good candidate. The $k$-DZ transformation is then applied a single time to this stable chaotic trajectory. Our claim is that this strategy quickly generates secure and light PRNs. Another strategy to generate secure PRNs, which might increase the computational cost to some extent, was proposed in Ref. \cite{saito2018pseudorandom}, where an approximate true trajectory of the Bernoulli map is calculated directly using real algebraic numbers.

\section*{Acknowledgments}

J. M. acknowledges a scholarship from the National Council for Scientific and Technological Development (CNPq grant \#155957/2018-0). 
O. M. B. acknowledges support from CNPq (grant \#307897/2018-4) and
FAPESP (grant \#16/18809-9).

\section*{Bibliography}
\nocite{*}
\bibliographystyle{apsrev4-2}
\bibliography{biblio-redu}

\begin{thebibliography}{20}%
\makeatletter
\providecommand \@ifxundefined [1]{%
 \@ifx{#1\undefined}
}%
\providecommand \@ifnum [1]{%
 \ifnum #1\expandafter \@firstoftwo
 \else \expandafter \@secondoftwo
 \fi
}%
\providecommand \@ifx [1]{%
 \ifx #1\expandafter \@firstoftwo
 \else \expandafter \@secondoftwo
 \fi
}%
\providecommand \natexlab [1]{#1}%
\providecommand \enquote  [1]{``#1''}%
\providecommand \bibnamefont  [1]{#1}%
\providecommand \bibfnamefont [1]{#1}%
\providecommand \citenamefont [1]{#1}%
\providecommand \href@noop [0]{\@secondoftwo}%
\providecommand \href [0]{\begingroup \@sanitize@url \@href}%
\providecommand \@href[1]{\@@startlink{#1}\@@href}%
\providecommand \@@href[1]{\endgroup#1\@@endlink}%
\providecommand \@sanitize@url [0]{\catcode `\\12\catcode `\$12\catcode
  `\&12\catcode `\#12\catcode `\^12\catcode `\_12\catcode `\%12\relax}%
\providecommand \@@startlink[1]{}%
\providecommand \@@endlink[0]{}%
\providecommand \url  [0]{\begingroup\@sanitize@url \@url }%
\providecommand \@url [1]{\endgroup\@href {#1}{\urlprefix }}%
\providecommand \urlprefix  [0]{URL }%
\providecommand \Eprint [0]{\href }%
\providecommand \doibase [0]{https://doi.org/}%
\providecommand \selectlanguage [0]{\@gobble}%
\providecommand \bibinfo  [0]{\@secondoftwo}%
\providecommand \bibfield  [0]{\@secondoftwo}%
\providecommand \translation [1]{[#1]}%
\providecommand \BibitemOpen [0]{}%
\providecommand \bibitemStop [0]{}%
\providecommand \bibitemNoStop [0]{.\EOS\space}%
\providecommand \EOS [0]{\spacefactor3000\relax}%
\providecommand \BibitemShut  [1]{\csname bibitem#1\endcsname}%
\let\auto@bib@innerbib\@empty
\bibitem [{\citenamefont {Slipantschuk}\ \emph {et~al.}(2013)\citenamefont
  {Slipantschuk}, \citenamefont {Bandtlow},\ and\ \citenamefont
  {Just}}]{slipantschuk2013relation}%
  \BibitemOpen
  \bibfield  {author} {\bibinfo {author} {\bibfnamefont {J.}~\bibnamefont
  {Slipantschuk}}, \bibinfo {author} {\bibfnamefont {O.~F.}\ \bibnamefont
  {Bandtlow}},\ and\ \bibinfo {author} {\bibfnamefont {W.}~\bibnamefont
  {Just}},\ }\href@noop {} {\bibfield  {journal} {\bibinfo  {journal} {Journal
  of Physics A: Mathematical and Theoretical}\ }\textbf {\bibinfo {volume}
  {46}},\ \bibinfo {pages} {75101} (\bibinfo {year} {2013})}\BibitemShut
  {NoStop}%
\bibitem [{\citenamefont {Pisarchik}\ and\ \citenamefont
  {Zanin}(2012)}]{pisarchik2012chaotic}%
  \BibitemOpen
  \bibfield  {author} {\bibinfo {author} {\bibfnamefont {A.~N.}\ \bibnamefont
  {Pisarchik}}\ and\ \bibinfo {author} {\bibfnamefont {M.}~\bibnamefont
  {Zanin}},\ }\href@noop {} {\bibfield  {journal} {\bibinfo  {journal}
  {International Journal of Computer Research}\ }\textbf {\bibinfo {volume}
  {19}},\ \bibinfo {pages} {49} (\bibinfo {year} {2012})}\BibitemShut {NoStop}%
\bibitem [{\citenamefont {Fridrich}(1997)}]{fridrich1997image}%
  \BibitemOpen
  \bibfield  {author} {\bibinfo {author} {\bibfnamefont {J.}~\bibnamefont
  {Fridrich}},\ }in\ \href@noop {} {\emph {\bibinfo {booktitle} {1997 IEEE
  International Conference on Systems, Man, and Cybernetics. Computational
  Cybernetics and Simulation}}},\ Vol.~\bibinfo {volume} {2}\ (\bibinfo
  {organization} {IEEE},\ \bibinfo {year} {1997})\ pp.\ \bibinfo {pages}
  {1105--1110}\BibitemShut {NoStop}%
\bibitem [{\citenamefont {Farajallah}\ \emph {et~al.}(2016)\citenamefont
  {Farajallah}, \citenamefont {El~Assad},\ and\ \citenamefont
  {Deforges}}]{farajallah2016fast}%
  \BibitemOpen
  \bibfield  {author} {\bibinfo {author} {\bibfnamefont {M.}~\bibnamefont
  {Farajallah}}, \bibinfo {author} {\bibfnamefont {S.}~\bibnamefont
  {El~Assad}},\ and\ \bibinfo {author} {\bibfnamefont {O.}~\bibnamefont
  {Deforges}},\ }\href@noop {} {\bibfield  {journal} {\bibinfo  {journal}
  {International Journal of Bifurcation and Chaos}\ }\textbf {\bibinfo {volume}
  {26}},\ \bibinfo {pages} {1650021} (\bibinfo {year} {2016})}\BibitemShut
  {NoStop}%
\bibitem [{\citenamefont {Zhang}\ \emph {et~al.}(2013)\citenamefont {Zhang},
  \citenamefont {Wong}, \citenamefont {Yu},\ and\ \citenamefont
  {Zhu}}]{zhang2013image}%
  \BibitemOpen
  \bibfield  {author} {\bibinfo {author} {\bibfnamefont {W.}~\bibnamefont
  {Zhang}}, \bibinfo {author} {\bibfnamefont {K.-W.}\ \bibnamefont {Wong}},
  \bibinfo {author} {\bibfnamefont {H.}~\bibnamefont {Yu}},\ and\ \bibinfo
  {author} {\bibfnamefont {Z.-L.}\ \bibnamefont {Zhu}},\ }\href@noop {}
  {\bibfield  {journal} {\bibinfo  {journal} {Communications in Nonlinear
  Science and Numerical Simulation}\ }\textbf {\bibinfo {volume} {18}},\
  \bibinfo {pages} {2066} (\bibinfo {year} {2013})}\BibitemShut {NoStop}%
\bibitem [{\citenamefont {Garasym}\ \emph {et~al.}(2016)\citenamefont
  {Garasym}, \citenamefont {Taralova},\ and\ \citenamefont
  {Lozi}}]{garasym2016new}%
  \BibitemOpen
  \bibfield  {author} {\bibinfo {author} {\bibfnamefont {O.}~\bibnamefont
  {Garasym}}, \bibinfo {author} {\bibfnamefont {I.}~\bibnamefont {Taralova}},\
  and\ \bibinfo {author} {\bibfnamefont {R.}~\bibnamefont {Lozi}},\ }in\
  \href@noop {} {\emph {\bibinfo {booktitle} {Complex Systems and Networks}}}\
  (\bibinfo  {publisher} {Springer},\ \bibinfo {year} {2016})\ pp.\ \bibinfo
  {pages} {131--161}\BibitemShut {NoStop}%
\bibitem [{\citenamefont {Vidal}\ \emph {et~al.}(2014)\citenamefont {Vidal},
  \citenamefont {Baptista},\ and\ \citenamefont {Mancini}}]{vidal2014fast}%
  \BibitemOpen
  \bibfield  {author} {\bibinfo {author} {\bibfnamefont {G.}~\bibnamefont
  {Vidal}}, \bibinfo {author} {\bibfnamefont {M.~S.}\ \bibnamefont
  {Baptista}},\ and\ \bibinfo {author} {\bibfnamefont {H.}~\bibnamefont
  {Mancini}},\ }\href@noop {} {\bibfield  {journal} {\bibinfo  {journal} {The
  European Physical Journal Special Topics}\ }\textbf {\bibinfo {volume}
  {223}},\ \bibinfo {pages} {1601} (\bibinfo {year} {2014})}\BibitemShut
  {NoStop}%
\bibitem [{\citenamefont {Cassanya}(2017)}]{cassanya2017method}%
  \BibitemOpen
  \bibfield  {author} {\bibinfo {author} {\bibfnamefont {G.~V.}\ \bibnamefont
  {Cassanya}},\ }\href@noop {} {\bibinfo {title} {Method for generating a
  pseudorandom sequence, and method for coding or decoding a data stream}}
  (\bibinfo {year} {2017}),\ \bibinfo {note} {{US Patent
  9,654,289}}\BibitemShut {NoStop}%
\bibitem [{\citenamefont {Lee}\ \emph {et~al.}(2003)\citenamefont {Lee},
  \citenamefont {Pei},\ and\ \citenamefont {Chen}}]{lee2003generating}%
  \BibitemOpen
  \bibfield  {author} {\bibinfo {author} {\bibfnamefont {P.-H.}\ \bibnamefont
  {Lee}}, \bibinfo {author} {\bibfnamefont {S.-C.}\ \bibnamefont {Pei}},\ and\
  \bibinfo {author} {\bibfnamefont {Y.-Y.}\ \bibnamefont {Chen}},\ }\href@noop
  {} {\bibfield  {journal} {\bibinfo  {journal} {Chinese Journal of physics}\
  }\textbf {\bibinfo {volume} {41}},\ \bibinfo {pages} {559} (\bibinfo {year}
  {2003})}\BibitemShut {NoStop}%
\bibitem [{\citenamefont {Machicao}\ and\ \citenamefont
  {Bruno}(2017)}]{machicao2017}%
  \BibitemOpen
  \bibfield  {author} {\bibinfo {author} {\bibfnamefont {J.}~\bibnamefont
  {Machicao}}\ and\ \bibinfo {author} {\bibfnamefont {O.~M.}\ \bibnamefont
  {Bruno}},\ }\href@noop {} {\bibfield  {journal} {\bibinfo  {journal} {Chaos:
  an interdisciplinary journal of nonlinear science}\ }\textbf {\bibinfo
  {volume} {27}},\ \bibinfo {pages} {53116} (\bibinfo {year}
  {2017})}\BibitemShut {NoStop}%
\bibitem [{\citenamefont {Graham}\ \emph {et~al.}(1989)\citenamefont {Graham},
  \citenamefont {Knuth}, \citenamefont {Patashnik},\ and\ \citenamefont
  {Liu}}]{graham1989concrete}%
  \BibitemOpen
  \bibfield  {author} {\bibinfo {author} {\bibfnamefont {R.~L.}\ \bibnamefont
  {Graham}}, \bibinfo {author} {\bibfnamefont {D.~E.}\ \bibnamefont {Knuth}},
  \bibinfo {author} {\bibfnamefont {O.}~\bibnamefont {Patashnik}},\ and\
  \bibinfo {author} {\bibfnamefont {S.}~\bibnamefont {Liu}},\ }\href@noop {}
  {\bibfield  {journal} {\bibinfo  {journal} {Computers in Physics}\ }\textbf
  {\bibinfo {volume} {3}},\ \bibinfo {pages} {106} (\bibinfo {year}
  {1989})}\BibitemShut {NoStop}%
\bibitem [{\citenamefont {Hilborn}\ \emph {et~al.}(2000)\citenamefont {Hilborn}
  \emph {et~al.}}]{hilborn2000chaos}%
  \BibitemOpen
  \bibfield  {author} {\bibinfo {author} {\bibfnamefont {R.~C.}\ \bibnamefont
  {Hilborn}} \emph {et~al.},\ }\href@noop {} {\emph {\bibinfo {title} {Chaos
  and nonlinear dynamics: an introduction for scientists and engineers}}}\
  (\bibinfo  {publisher} {Oxford University Press on Demand},\ \bibinfo {year}
  {2000})\BibitemShut {NoStop}%
\bibitem [{\citenamefont {Saito}\ and\ \citenamefont
  {Yamaguchi}(2018)}]{saito2018pseudorandom}%
  \BibitemOpen
  \bibfield  {author} {\bibinfo {author} {\bibfnamefont {A.}~\bibnamefont
  {Saito}}\ and\ \bibinfo {author} {\bibfnamefont {A.}~\bibnamefont
  {Yamaguchi}},\ }\href@noop {} {\bibfield  {journal} {\bibinfo  {journal}
  {Chaos: An Interdisciplinary Journal of Nonlinear Science}\ }\textbf
  {\bibinfo {volume} {28}},\ \bibinfo {pages} {103122} (\bibinfo {year}
  {2018})}\BibitemShut {NoStop}%
\bibitem [{\citenamefont {May}(1976)}]{MayChaos}%
  \BibitemOpen
  \bibfield  {author} {\bibinfo {author} {\bibfnamefont {R.~M.}\ \bibnamefont
  {May}},\ }\href@noop {} {\bibfield  {journal} {\bibinfo  {journal} {Nature}\
  }\textbf {\bibinfo {volume} {261}},\ \bibinfo {pages} {459} (\bibinfo {year}
  {1976})}\BibitemShut {NoStop}%
\bibitem [{\citenamefont {Ott}(2002)}]{OttChaosBook}%
  \BibitemOpen
  \bibfield  {author} {\bibinfo {author} {\bibfnamefont {E.}~\bibnamefont
  {Ott}},\ }\href@noop {} {\emph {\bibinfo {title} {Chaos in Dynamical
  Systems}}}\ (\bibinfo  {publisher} {Cambridge University Press},\ \bibinfo
  {year} {2002})\BibitemShut {NoStop}%
\bibitem [{\citenamefont {Grebogi}\ \emph {et~al.}(1990)\citenamefont
  {Grebogi}, \citenamefont {Hammel}, \citenamefont {Yorke},\ and\ \citenamefont
  {Sauer}}]{grebogi1990shadowing}%
  \BibitemOpen
  \bibfield  {author} {\bibinfo {author} {\bibfnamefont {C.}~\bibnamefont
  {Grebogi}}, \bibinfo {author} {\bibfnamefont {S.~M.}\ \bibnamefont {Hammel}},
  \bibinfo {author} {\bibfnamefont {J.~A.}\ \bibnamefont {Yorke}},\ and\
  \bibinfo {author} {\bibfnamefont {T.}~\bibnamefont {Sauer}},\ }\href@noop {}
  {\bibfield  {journal} {\bibinfo  {journal} {Physical Review Letters}\
  }\textbf {\bibinfo {volume} {65}},\ \bibinfo {pages} {1527} (\bibinfo {year}
  {1990})}\BibitemShut {NoStop}%
\bibitem [{\citenamefont {Alligood}\ \emph {et~al.}(2012)\citenamefont
  {Alligood}, \citenamefont {Sauer},\ and\ \citenamefont
  {Yorke}}]{AlligoodBookChaos}%
  \BibitemOpen
  \bibfield  {author} {\bibinfo {author} {\bibfnamefont {K.}~\bibnamefont
  {Alligood}}, \bibinfo {author} {\bibfnamefont {T.}~\bibnamefont {Sauer}},\
  and\ \bibinfo {author} {\bibfnamefont {J.}~\bibnamefont {Yorke}},\
  }\href@noop {} {\emph {\bibinfo {title} {Chaos: An Introduction to Dynamical
  Systems}}},\ Textbooks in Mathematical Sciences\ (\bibinfo  {publisher}
  {Springer New York},\ \bibinfo {year} {2012})\BibitemShut {NoStop}%
\bibitem [{\citenamefont {Boyd}\ \emph {et~al.}(2009)\citenamefont {Boyd},
  \citenamefont {Diaconis}, \citenamefont {Parrilo},\ and\ \citenamefont
  {Xiao}}]{boyd2009fastest}%
  \BibitemOpen
  \bibfield  {author} {\bibinfo {author} {\bibfnamefont {S.}~\bibnamefont
  {Boyd}}, \bibinfo {author} {\bibfnamefont {P.}~\bibnamefont {Diaconis}},
  \bibinfo {author} {\bibfnamefont {P.}~\bibnamefont {Parrilo}},\ and\ \bibinfo
  {author} {\bibfnamefont {L.}~\bibnamefont {Xiao}},\ }\href@noop {} {\bibfield
   {journal} {\bibinfo  {journal} {SIAM Journal on Optimization}\ }\textbf
  {\bibinfo {volume} {20}},\ \bibinfo {pages} {792} (\bibinfo {year}
  {2009})}\BibitemShut {NoStop}%
\bibitem [{\citenamefont {Marsaglia}(1998)}]{DIEHARD}%
  \BibitemOpen
  \bibfield  {author} {\bibinfo {author} {\bibfnamefont {G.}~\bibnamefont
  {Marsaglia}},\ }\href {http://www.stat.fsu.edu/pub/diehard} {\bibinfo {title}
  {The {M}arsaglia random number {{CDROM}}, with the {DIEHARD} battery of tests
  of randomness}} (\bibinfo {year} {1998})\BibitemShut {NoStop}%
\bibitem [{\citenamefont {Rukhin}\ \emph {et~al.}(2001)\citenamefont {Rukhin},
  \citenamefont {Soto}, \citenamefont {Nechvatal}, \citenamefont {Smid},
  \citenamefont {Barker}, \citenamefont {Leigh}, \citenamefont {Levenson},
  \citenamefont {Vangel}, \citenamefont {Banks},\ and\ \citenamefont
  {Heckert}}]{NIST-PRNG}%
  \BibitemOpen
  \bibfield  {author} {\bibinfo {author} {\bibfnamefont {A.}~\bibnamefont
  {Rukhin}}, \bibinfo {author} {\bibfnamefont {J.}~\bibnamefont {Soto}},
  \bibinfo {author} {\bibfnamefont {J.}~\bibnamefont {Nechvatal}}, \bibinfo
  {author} {\bibfnamefont {M.}~\bibnamefont {Smid}}, \bibinfo {author}
  {\bibfnamefont {E.}~\bibnamefont {Barker}}, \bibinfo {author} {\bibfnamefont
  {S.}~\bibnamefont {Leigh}}, \bibinfo {author} {\bibfnamefont
  {M.}~\bibnamefont {Levenson}}, \bibinfo {author} {\bibfnamefont
  {M.}~\bibnamefont {Vangel}}, \bibinfo {author} {\bibfnamefont
  {D.}~\bibnamefont {Banks}},\ and\ \bibinfo {author} {\bibfnamefont
  {A.}~\bibnamefont {Heckert}},\ }\href@noop {} {\emph {\bibinfo {title} {{NIST
  Special Publication 800-22: A statistical test suite for random number
  generator for criptographic applications}}}},\ \bibinfo {type} {Tech. Rep.}\
  (\bibinfo  {institution} {National Institute of Standards and Technology},\
  \bibinfo {address} {Gaithersburg, MD, USA},\ \bibinfo {year}
  {2001})\BibitemShut {NoStop}%
\end{thebibliography}%

\end{document}